\documentclass[12pt]{iopart}
\usepackage{fancyhdr}
\usepackage{graphicx}
\usepackage{float}
\usepackage{comment}
\usepackage{subcaption}
\usepackage[font=small,labelfont=bf]{caption}

\expandafter\let\csname equation*\endcsname\relax
\expandafter\let\csname endequation*\endcsname\relax
\usepackage{amsmath}
\usepackage{todonotes}
\usepackage{amsfonts}
\usepackage{amssymb}
\usepackage{amsthm}
\usepackage{bm}
\usepackage{listings}
\usepackage{pdfpages}
\usepackage{appendix}

\usepackage[backend = biber,style = apa,block=ragged]{biblatex}
\begin{document}
\title[Quantifying risk of a noise-induced AMOC collapse]{Quantifying risk of a noise-induced AMOC collapse from northern and tropical Atlantic Ocean variability}
\author{R Chapman$^{1,*}$, P Ashwin $^1$, J Baker$^2$ and R A Wood$^2$ 
}
\address{$^1$ Department of Mathematics and Statistics, University of Exeter, Exeter EX8 4QF, UK}
\address{$^2$ Met Office Hadley Centre, Exeter, U.K.}
\address{$^*$ Corresponding author: rc686@exeter.ac.uk and ruth.chapman@nbi.ku.dk}

\begin{abstract}

The Atlantic Meridional Overturning Circulation (AMOC) exerts a major influence on global climate. There is much debate about whether the current strong AMOC  may collapse as a result of anthropogenic forcing and/or internal variability. Increasing the noise in simple salt-advection models can change the apparent AMOC tipping threshold. However, it's not clear if `present-day' variability is strong enough to induce a collapse.
Here, we investigate how internal variability affects the likelihood of AMOC collapse. We examine the internal variability of basin-scale salinities and temperatures in four CMIP6 pre-industrial simulations. We fit this to an empirical, process-based AMOC box model, and
find that noise-induced AMOC collapse (defined as a decade in which the mean AMOC strength falls below $5 \ \rm{Sv}$) is unlikely, however, if the AMOC is pushed closer to a bifurcation point due to external climate forcing, noise-induced tipping becomes more likely.
Surprisingly, we find a case where forcing temporarily overshoots a stability threshold but noise decreases the probability of collapse. Accurately modelling internal decadal variability is essential for understanding the increased uncertainty in AMOC projections.
\end{abstract}

\noindent{\it  AMOC, Tipping Points, Stochastic Dynamics} 

\maketitle

\section{Introduction}

The Atlantic Meridional Overturning Circulation (AMOC) is an important ocean circulation that is responsible for bringing warm water north towards Western Europe, contributing to the temperate climate at these high latitudes. The greater insolation and evaporation into the atmosphere in the tropical and sub-tropical regions makes the surface water saltier and warmer. This water flows northwards into the North Atlantic where it cools, becomes denser and therefore sinks. This meridional overturning depends on a combination of salinity- and temperature-driven buoyancy effects. A variety of ocean models (from simple models to fully coupled Global Circulation Model (GCM) simulations) suggest there can be bi-stability between the current `on' state and a much weaker `off' state \parencite{dijkstra05, stommel61, Jackson15, Romanou2023,VanWesten2024}, with the so-called salt-advection feedback playing a key role in the transition between states \parencite{stommel61,Jackson2017,Weijer19}. Indeed, paleo-data suggests that in the past this circulation has transitioned into the `off' state \parencite{Bryan86, Monahan08,Weijer19, Ritz13}. Some observational studies suggest a recent weakening of the AMOC \parencite{Jackson2022, Rahmstorf2015, Smeed2018}, and a recent study suggested that a collapse is possible this century \parencite{Ditlevsen2023}. Nonetheless, the most recent IPCC report \parencite{IPCC_AR6} and estimates from the latest Coupled Model Inter-comparison Project, Phase 6 (CMIP6) classify this as unlikely \parencite{Baker2023,Weijer2020}.

Previous studies have identified three main mechanisms for tipping in dynamical systems models; bifurcation-, rate- and noise-induced tipping \parencite{ashwin12}. 
Past studies have shown that adding noise to a simple salt-advection feedback model \parencite{stommel61} can cause noise-induced tipping \parencite{Monahan08}.
\textcite{Castellana2019} estimate noise forcing from observational data relative to the freshwater forcing on annual time scales. When using this as a lower bound for noise forcing in a simple box model \parencite{Cimatoribus2014}, they find transition probabilities for the temporary collapse of the overturning are only non-negligible for some higher noise and sufficiently bi-stable AMOC. For a persistent collapse, the probabilities are very low in the time period considered. 

Studies of North Atlantic climate variability have typically separated out noise terms from shorter, inter-annual time-scale atmospheric forcing (such as the North Atlantic Oscillation (NAO) \parencite{Mecking2014}) and freshwater forcing \parencite{Menary2015}. Other studies \parencite{Martin2019} have examined inter-decadal variability in the North Atlantic and find that mechanisms differ between GCMs, with temperature driving variability in one class of models, while another class is dominated by salinity. 

In this paper, we determine whether internal climate `noise' in the North Atlantic can substantially influence the likelihood of AMOC collapse, which we have defined here as a decade in which the decadal-mean AMOC strength falls below $5 \ \rm{Sv}$ \parencite{Jackson2023}.
We consider decadal time series of salinity, temperature and AMOC strength, which are more likely to influence the basin-scale AMOC than higher-frequency variations which have a more regional influence \parencite{Jackson2022}.
It is difficult to assess scenarios such that a `real-world' noise forcing can cause noise-tipping in GCMs, given the length of runs required. We, therefore, use a simple process-based AMOC box model \parencite{wood19,Alkhayuon19} which is known to exhibit bifurcation- and rate-induced tipping, and add a `noise' term to represent stochastic variability. This stochastic or noise term physically represents the drivers of natural variability on basin length scales and decadal time scales. This includes different types of box mixing, via ocean or atmospheric processes.
That is, the noise captures the drivers of internal variability for processes in the GCM that produce perturbations from equilibrium salinities in the deterministic system.

Here we go beyond the CMIP5 results of \textcite{Castellana2019} by fitting noise from a selection of CMIP6 climate models for which suitable data were available (Table S1). We extract fluctuations of volume-averaged salinity and temperature and calibrate the noise term in our simple box model to fit these and diagnose the characteristics of the noise. 
The box model used represents key ocean processes that impact salinity changes, considering the entire depth of the North Atlantic region, and has model parameters calibrated to modern GCMs. More complex processes which are not included in the deterministic system, but are present in the GCMs, are captured in the stochastic term.

We assume that the ‘forcing’ noise originates from processes independent of the box model, in particular ocean and atmospheric turbulence, precipitation, and ice sheet melt, which are then filtered through the complex dynamics of the CMIP GCMs to produce variability on the space- and time-scales that are resolved by the box model (basin and decadal). Note also that by sampling decadal data from the GCMs we are sampling the outcome of a (nonlinear) Hasselmann process (modified by the complex GCM dynamics) rather than the forcing directly, and the likelihood estimates will be agnostic to the form or source of the inputs.
We then apply this noise alongside various forcing scenarios, which can represent climate change impacts via ice sheet melt, to examine if the estimated noise can substantially alter the tipping probability. Because there is evidence that modern GCMs underestimate `present-day' variability \parencite{Menary2015,Romanou2023,Roberts2020}, we also consider some larger noise forcing while maintaining the covariance properties between regions. 

The paper is organized as follows: in Section~\ref{sec:variability} we discuss variability of AMOC strength and regional salinity and temperature in four CMIP6 model pre-industrial control (piControl) experiments. 
In Section \ref{sec:nonlinear}, we calibrate our simple stochastic AMOC model to the CMIP6 pre-industrial control runs.
In Section \ref{sec:tipping_probabilities}, we use the calibrated models to calculate probabilities of AMOC collapse over a 1000 year period with different forcing and noise. Our conclusions are presented in Section \ref{sec:conclusions}.

\section{Atlantic Ocean variability in CMIP6 models}
\label{sec:variability}

\subsection{Variability of AMOC strength}
\label{sec:amoc_variability}

\begin{figure}
    \centering
    \includegraphics[width = 0.5\linewidth]{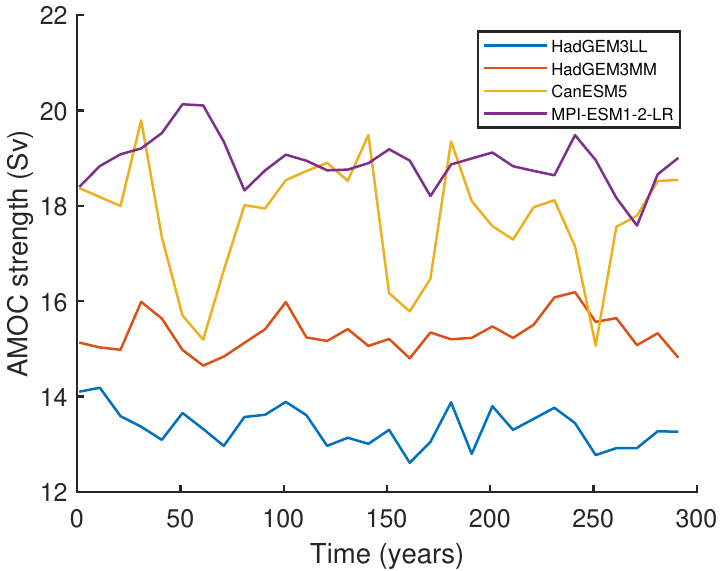}
    \caption{Time-series of decadal averaged AMOC strength, $q$, for the CMIP6 models considered at $30^{\circ} S$.}
    \label{fig:q_timeseries}
\end{figure}

Figure \ref{fig:q_timeseries} shows the decadal averaged time series of $q$, the strength of the AMOC at $30^{\circ}S$, for the CMIP6 models considered. We consider {\em HadGEM3-GC3.1-LL} (abbreviated as {\tt HadGEM3LL}), {\em HadGEM3-GC3.1-MM} (abbreviated as {\tt HadGEM3MM}), {\em CanESM5} and {\em MPI-ESM1-2-LR} (abbreviated as {\tt MPI}), details being listed in Table S1. For all of the pre-industrial control runs 
extracted, the AMOC remains in an `on' state, which corresponds to the present-day strong circulation. The models were selected from the subset of those used in \textcite{Jackson2023}, based on availability of the required data for this study.
We choose to use decadal time series since this variability would have the biggest impact on basin scales and the dynamics of large-scale AMOC tipping. 

Table S2 lists some basic statistics, in particular the time-mean of the AMOC, $\overline{q}$, and the variance of AMOC strength at $30 ^{\circ} S$, $\Sigma(q)$. The AMOC at $30 ^{\circ} S$ was used because our model focuses on basin-scale AMOC variations, which are likely to have the largest climatic impact on decadal timescales \parencite{wood19}.
We find a significant difference in the decadal variability of $q$ from these timeseries. 
While most models considered have a small variance ($0.14-0.27 \ (\rm{Sv})^2$), {\tt CanESM5} has a significantly higher variance of $1.49 \ (\rm{Sv})^2$, which can be seen in Figure~ \ref{fig:q_timeseries}. This large difference highlights inconsistencies between CMIP6 models.

There is, however, evidence that all CMIP6 models systematically underestimate the variability in the ocean \parencite{Roberts2020}. Observational results \parencite{Smeed2018} and historical hydrographic timeseries \parencite{Fraser2021} show significantly higher variability in both decadal and annual timeseries.
For example, {\tt HadGEM3MM} has a decadal variance of $0.15 \ (\rm{Sv})^2$ and an annual variance of $0.71 \ (\rm{Sv})^2$ in the piControl experiment. 
This is compared to an annual variance for the RAPID array of $3.0 \ (\rm{Sv})^2$ and the variability of amplitude of a reconstructed AMOC time series from historical observations of $\sim 2 \ \rm{Sv}$ \parencite{Fraser2021} which corresponds to a variance of $4 \ (\rm{Sv})^2$.

We first consider the risk of an AMOC collapse using the variance of $q$ found in the four CMIP6 piControl simulations examined. We use $5 \ \rm{Sv}$ as a limit for collapse as in \textcite{Jackson2023}- that is we consider a run to have collapsed if the decadal mean AMOC strength drops below this threshold. We compare the standard deviation of the decadal-mean AMOC strength from the piControl runs to the average value to estimate how many standard deviations from $q = 5 \ \rm{Sv}$ the mean state is. This allows us to determine the probability of noise-induced AMOC collapse in a linear model with no external forcing. Here, we present the result for {\tt HadGEM3MM}.
We find $\bar{q} = 15.3115 \ \rm{Sv}$, variance $\Sigma = 0.1497 \ (\rm{Sv})^2$, therefore the standard deviation is $\sigma = 0.3869 \ \rm{Sv}$ which corresponds to the mean state being over $26$ standard deviations from $5 \ \rm{Sv}$. This corresponds to a transition probability of effectively zero over the Earth's lifetime. 

Some preliminary work using linear models showed that a salinity-driven model is a good fit for these time series: adding temperature dependence does not improve the result. These results are not included in the scope of this paper. 

The levels of AMOC variability diagnosed from the CMIP6 models lead to vanishingly small likelihood of an AMOC collapse if the AMOC variability were simply a Gaussian process (by our definition of collapse requiring the AMOC strength to drop below $5 \ \rm{Sv}$ for over a decade). However, since the AMOC may be in a bi-stable regime, the distribution may be non-Gaussian (or even bimodal) over long enough timescales. This motivates our study of the impact of noise in a process-based nonlinear AMOC box model in Section \ref{sec:3box_estimates}.

\subsection{Variability of Atlantic salinity}
\label{sec:var_salinity}

\begin{figure}
    \centering
    \includegraphics[width=10cm]{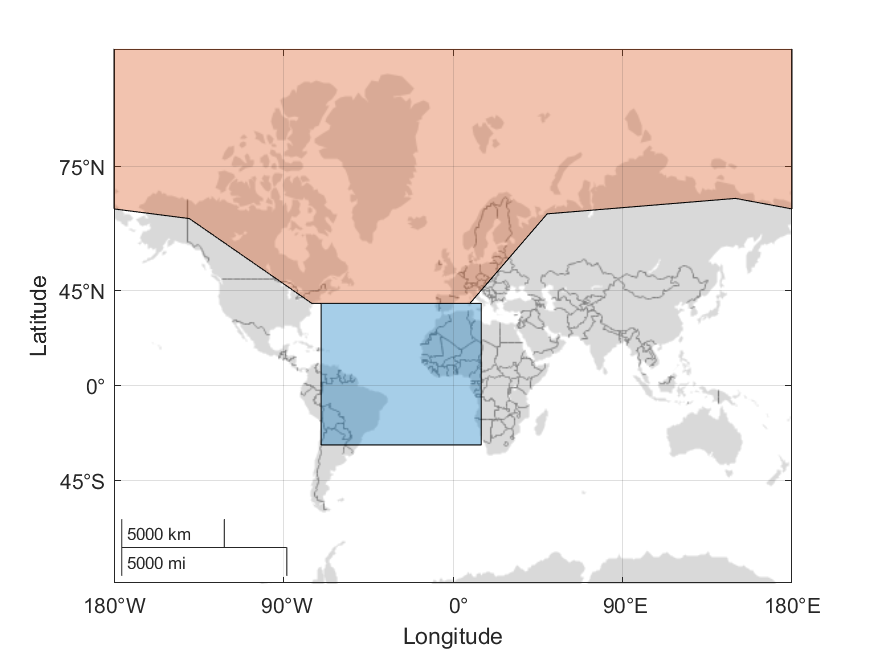}
   
    \caption{Projection of the Northern and Tropical boxes on the surface. The orange Northern box (Nor. box) is defined as the Atlantic and Arctic Ocean North of $40^{\circ} \rm{N}$ at all depths. The blue Tropical box (Trop. box) is above the Atlantic thermocline, for latitudes between $30 ^{\circ} \rm{S}$ and $40 ^{\circ} \rm{N}$.}
    \label{fig:boxregions}
\end{figure}

To quantify the variability of the AMOC for the box model, we examine the regional volume-averaged salinities and temperatures of two regions in the Atlantic Ocean. 
The Northern region (Nor. box) is defined as the Atlantic and Arctic Oceans north of $40^{\circ} \rm{N}$, at all depths. The Tropical thermocline region (Trop. box) is defined as the upper $995 \ \rm{m}$ of the Atlantic Ocean from $40 ^{\circ} \rm{N}$ to around $30 ^{\circ} \rm{S}$. The surface outline of the boxes is shown in Figure~\ref{fig:boxregions}.
These volumes are the basis of the box model of \textcite{wood19}, and its subsequent simplification to three boxes by \textcite{Alkhayuon19},  which is used in Section \ref{sec:nonlinear}.  The boundaries were chosen based on the zonal-average salinity distributions, which allow us to separate the large-scale water masses, following Figure 1b of \textcite{wood19}.
For each model, $300$ years of decadal volume-averaged data were extracted from piControl experiments.

\begin{figure}
    \centering
    \includegraphics[width = 1\linewidth]{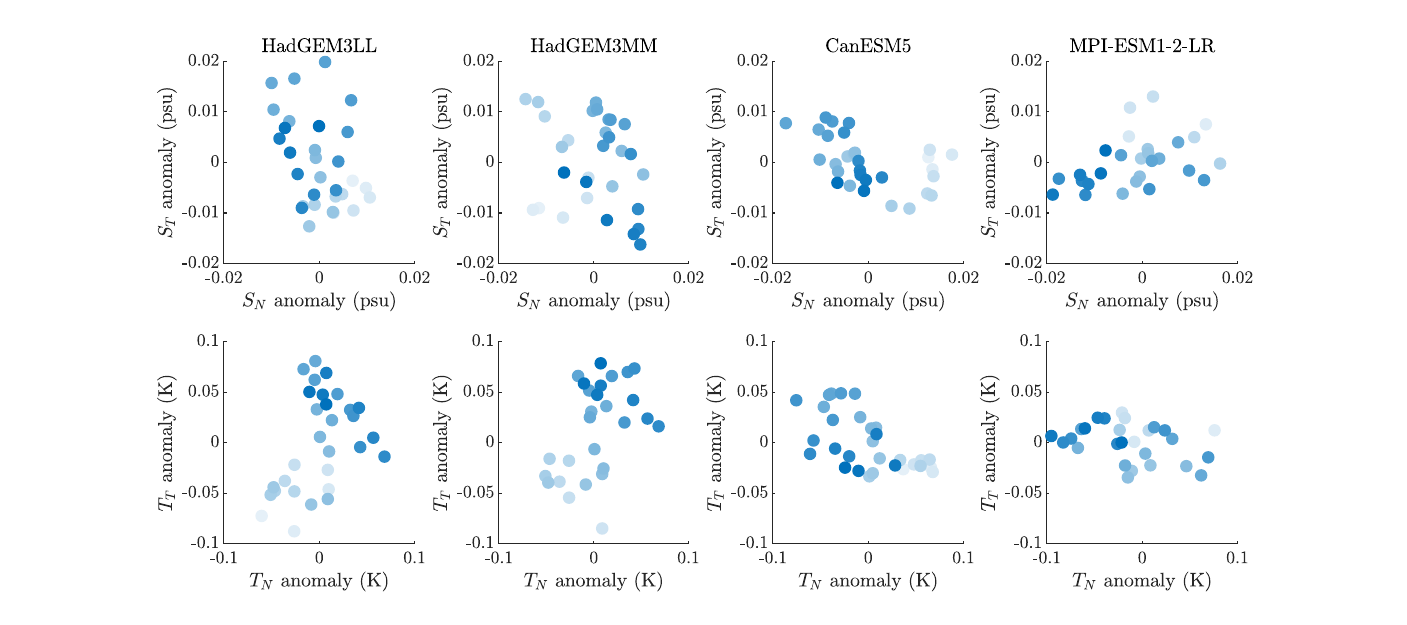}
    \caption{Anomalies of decadal averages of salinity and temperature for the CMIP6 piControl model runs, and boxes discussed. Each panel shows a phase space of salinity (top row, temperature on bottom row) and plots the anomaly from the mean of the decadal timeseries in the $\rm{Nor.}$ and $\rm{Trop.}$ boxes (labelled $N$ and $T$ respectively). Each column shows a different CMIP6 model. Plotted as a scatter, colour depth increases with time. }
    \label{fig:Salinity_temp_data} 
\end{figure}

Figure \ref{fig:Salinity_temp_data} shows the decadal average salinity and temperature time series for each CMIP6 run and each box, plotted as anomalies from the mean in a phase portrait. Summary statistics of these timeseries are shown in Table S2. 

We see a generally consistent amplitude of variability across models and variables. 
Considering the variance in salinity and temperature and the estimates for $\alpha,\beta$ in the equation of state for AMOC strength (Equation \ref{eq:q}), we find that salinity variability dominates over temperature variability in the contribution to density (and hence AMOC) variability.

Some models have noticeably higher variance in one variable, for example {\tt MPI} has significantly more variance in $S_{\rm{Nor.}}$ than $S_{\rm{Trop.}}$. We see similar results in the temperature plots in the second row of Figure ~\ref{fig:Salinity_temp_data}, with {\tt HadGEM3} models having higher temperature variance in $T_{\rm{Trop.}}$, and {\tt MPI} having a much higher variance in $T_{\rm{Nor.}}$ than $T_{\rm{Trop.}}$.
\section{Fitting to a data-adapted nonlinear three-box model}
\label{sec:nonlinear}

We now fit our regional salinity data to the three-box model of \textcite{Alkhayuon19} which has a bi-stable AMOC. 
This model is reduced from a five-box model of \textcite{wood19}, which comprises boxes in the North Atlantic (Nor.), Tropical thermocline (Trop.), Southern Ocean (S), Deep Water (B) and Indo-Pacific thermocline (IP). The Northern and Tropical boxes correspond to the ocean volumes already considered and shown in Figure \ref{fig:boxregions}.

It is possible to make the reduction to three boxes by assuming that two boxes have constant salinity (here the `S' and `B' boxes), inferred from the calibration to {\tt FAMOUS} in \textcite{wood19}. Due to the conservation of total salt, one of the three non-dynamic boxes (here the Indo-Pacific box) can be solved to give a two-dimensional system for the northern and tropical salinities.

The boxes interact by allowing the transfer of salt as in \textcite{stommel61}. We also account for freshwater fluxes into the surface boxes, and for wind-driven gyre transports between surface boxes. The circulation overturns in the North Atlantic box and is allowed to upwell in either the Southern Ocean or the Indo-Pacific.
The AMOC strength, $q$, is given by;
\begin{equation}
    q = \lambda ( \alpha (T_{\rm{S}} - T_{\rm{Nor.}}) + \beta (S_{\rm{Nor.}} - S_{\rm{S}})),
    \label{eq:q}
\end{equation}
where $\alpha$ and $\beta$ represent a linearised equation of state for seawater, and $\lambda$ is a model parameter defined by the calibration.
$T$ is a box-averaged temperature and $S$ is a box-averaged salinity. The subscripts indicate the box label.
We determine the AMOC state based on the sign of $q$; $q>0$ corresponds to an `on' state, $q<0$ corresponds to an `off' state and a reversal of the circulation. 

Previous work using this box model used parameters, representing the large-scale AMOC dynamics, that had been calibrated to the {\tt FAMOUS} GCM \parencite{Jones2005}. In this work, we have re-calibrated the five-box model to {\tt HadGEM3LL} and {\tt HadGEM3MM}, which is advantageous since these are more recent GCMs of higher resolution, and therefore calibration to the three-box model is expected to be closer to `real-world'. The full equations for the three-box model and the updated model parameters are presented in Supplementary Information, S3, with a small optimisation made to the values of the freshwater fluxes in Section \ref{sec:3box_estimates}, to account for the reduction to three boxes. 
The necessary data to calibrate the box model parameters to {\tt CanESM5} and {\tt MPI} were not available; therefore we use the {\tt HadGEM3MM} calibration of the box model parameters as the default, and apply noisy profiles from {\tt HadGEM3MM}, {\tt CanESM5} and {\tt MPI} (see below) to this version. For {\tt HadGEM3LL} we calibrate both the box model parameters themselves and the noise parameters to the underlying {\tt HadGEM3LL} GCM.

In order to consider noise-induced tipping, we include stochastic terms with non-trivial covariance. We write the model as
\begin{equation}
d\bm{x}(t) = \bm{f}(\bm{x}(t),t) dt + \bm{B}\, d\bm{W}_t,
\label{eq:nonlinear+noise}
\end{equation}
where $\bm{x}(t)=(S_{\rm{Nor.}},S_{\rm{Trop.}})$ is a vector that represents the salinities in the $\rm{Nor.}$ and $\rm{Trop.}$ regions of Figure \ref{fig:boxregions} at a time step $t$ and $\bm{f}$ is the box model from S3. We assume stationary additive white noise; $\bm{B}$ is a constant matrix of driving noise amplitudes and $\bm{W}_t$ is a vector of standard independent Wiener processes (each component of $\bm{W}_t$ is a Brownian motion with zero mean). 
Note that only if $\bm{B}$ is diagonal then there are independent noise processes acting on each component.

An approximate negative log-likelihood method was used \parencite{sarkka19} to fit the noise model to the decadal averages for each of the four CMIP6 control runs. 
More details are given in S3 and S4. 
In each case, 
the log-likelihood method was used to minimise the bias between the box model simulations and the salinity time series extracted for the CMIP6 models (Figure \ref{fig:Salinity_temp_data}). The log-likelihood method can be used to estimate a number of `unknown' noise parameters of the system. 
The estimated noise covariance and some optimised freshwater flux values are included in Table S4, and the noise covariance decomposed into the noise amplitudes, $\bm{B}$, presented in the following sub-section. This optimisation ensures we are only estimating the noise term in $\bm{B}$, and not some offset in the system or other dynamics.
This method was found to give a better fit of the box model to the GCM salinities than the method used by \textcite{Alkhayuon19}. 

\subsection{Results of parameter estimation for the nonlinear model from CMIP6 data}
\label{sec:3box_estimates}

We present optimal estimated noise amplitudes (Table \ref{tab:3boxestimated}) for the three-box model (Equation ~\ref{eq:nonlinear+noise}) for the various CMIP6 runs considered. 
Figure S1 shows a box model salinity time series with the optimisation for {\tt HadGEM3MM}, compared with the volume averaged GCM time series, demonstrating a good fit for these optimal parameters. 
\begin{table}
\centering
\begin{tabular}{l c c c c}
\hline
Model  & {\tt HadGEM3LL} & {\tt HadGEM3MM} & {\tt CanESM5} & {\tt MPI}  \\
\hline
$\bm{B}_{11} (\times 10^{-2})$ ($\rm{psu} \times (\rm{year})^{-\frac{1}{2}}$)&  0.3369 & 0.1263 & 0.3860 & 0.8686\\ 
$\bm{B}_{21}(\times 10^{-2})$ ($\rm{psu} \times (\rm{year})^{-\frac{1}{2}}$) &  -0.0551 & -0.0869 & -0.1579 & -0.0443\\ 
$\bm{B}_{22} (\times 10^{-2})$ ($\rm{psu} \times (\rm{year})^{-\frac{1}{2}}$)&  0.4006 & 0.1088 & 0.2792 & 0.3543\\ \hline
\end{tabular}
\caption{Estimated noise amplitudes for the nonlinear 3-box model (Equation~\ref{eq:nonlinear+noise}) estimated using the approximated  log likelihood function from S4 for the decadal CMIP6 data shown in Figure \ref{fig:Salinity_temp_data}. $\bm{B}$ is the noise amplitudes, where the subscripts indicate matrix component (1 = Nor., 2 = Trop.). Note that $\bm{B}_{12}=0$.}
\label{tab:3boxestimated}
\end{table}

From Table~\ref{tab:3boxestimated} we draw some conclusions about the estimated noise driving the box model.  
We see that the noise amplitude in the Northern box, $\bm{B}_{11}$, is generally higher than that in the Tropical box ($\bm{B}_{22}$). This is an expected result since there are many more processes happening in the North Atlantic, such as sea-ice interactions and deep water formation, which are likely to impact salinity. We also observe that there is still significant disagreement between models, for example, the noise amplitude in the $\rm{Nor.}$ box ($\bm{B}_{11}$) varies by a factor of four between models, suggesting little convergence of models from CMIP5 to CMIP6 \parencite{Castellana2019}.

Finally, we note that the driving noise amplitudes ($\bm{B}_{ij}$) in Table~\ref{tab:3boxestimated} are very small compared to e.g. the global salinity, meaning we do not expect a strong stochastic response in the box model. This confirms that the variability in the model is very small and that the driving noise alone cannot be expected to be enough to collapse the AMOC without additional freshwater forcing and/or more realistic variability.

\section{Probability of AMOC collapse} 
\label{sec:tipping_probabilities}

In this section, we calculate AMOC collapse probabilities for the calibrated three-box model under different scenarios of noise and freshwater forcing. As before, we define `collapse' to be at least one decade when the AMOC is below $5 \ \rm{Sv}$ \parencite{Jackson2023}. We calculate the collapse probabilities using a Monte Carlo simulation of 1000 realisations over 1000 years. 

\begin{figure}
    \centering
    \includegraphics[width = 0.7\linewidth]{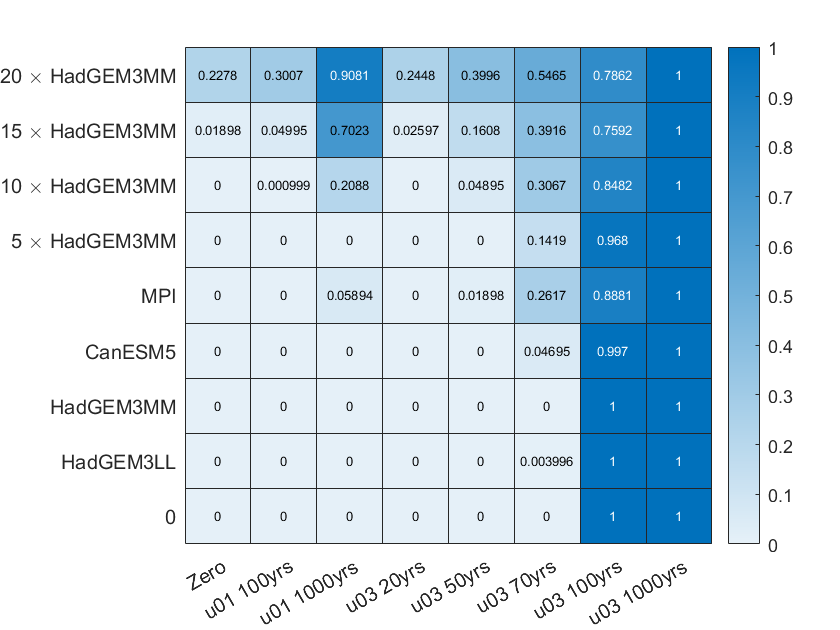}
    \caption{A heat map of collapse probabilities within the first 1000 years for different freshwater forcing functions ($x$-axis, ``Zero'' represents the pre-industrial control, other control scenarios as described in the text) and varying noise matrices ($y$-axis), run using the three-box model. 
    Noise amplitudes (rows) are: zero; the estimated decadal noise from {\tt HadGEM3LL, HadGEM3MM, CanESM5 and MPI}; and then 5,10,15 and 20 times the {\tt HadGEM3MM} noise amplitude matrix. In every case except {\tt HadGEM3LL}, the large scale box model parameters are set to {\tt HadGEM3MM} values (see text Section \ref{sec:nonlinear}).
    Each probability is calculated using Monte Carlo simulations of the box model with 1000 realisations of the noise for 1000 years.   
    }
    \label{fig:heatmap}
\end{figure}

Figure \ref{fig:heatmap} shows the resulting probabilities for a range of freshwater forcing scenarios and noise levels. In each experiment, the box model parameters have been calibrated to {\tt HadGEM3MM}, except {\tt HadGEM3LL} noise level which is calibrated to itself, and then the noise and freshwater correction in Table \ref{tab:3boxestimated} and Table S4 is applied for each model calibration.
The labels on the $y$-axis in Figure \ref{fig:heatmap} indicate the noise level used in each set of simulations, including no noise forcing, the CMIP6 model estimates from Section \ref{sec:3box_estimates}, and inflated {\tt HadGEM3MM} noise (the noise amplitudes are multiplied by a scalar, and maintain the covariance structure), since the GCMs may be underestimating the noise amplitude in the present day. 
We consider different profiles of North Atlantic freshwater forcing (hosing) as used in the recent NAHosMIP study \textcite{Jackson2023} along the $x$-axis. These are instantaneous hosing functions with uniform forcing strengths across the Northern box of $0.3 \ \rm{Sv}$ and $0.1 \ \rm{Sv}$ (labelled as $u01$ and $u03$ respectively) applied for different lengths of time, after which the forcing is returned to zero. These are labelled as e.g; $70\ yrs$ for $70$ years of forcing. This offers a good range of hosing scenarios since $0.3 \ \rm{Sv}$ is larger than the bifurcation point for the {\tt HadGEM3MM} calibration \parencite{Chapman2024}, while $0.1 \ \rm{Sv}$ is much smaller, allowing us to determine the impact of the noise on the probability of tipping, in cases where the forcing is above or below the critical level for bifurcation-induced tipping. These forcing profiles are designed to represent the top-end of climate forcing scenarios such as freshwater input from glacier melt in Greenland \parencite{Jackson2023}. Since we may also expect variance to increase under a forced AMOC (in line with the theory of critical slowing down \parencite{Boulton2014,Boers22,Lenton11}), the inflated noise combined with these hosing scenarios can be considered to be a plausible, if highly idealised, high-end storyline for the AMOC under climate change.

From the first column of Figure ~\ref{fig:heatmap}, we see that in the absence of freshwater forcing, an AMOC collapse is very unlikely unless the noise is substantially higher than the GCM estimates. Adding freshwater forcing to move the AMOC nearer to its bifurcation point results in an increase of the probability of collapse, especially for noise from the {\tt MPI} model, which has higher variability in $S_{\rm{N}}$ than the other models (Figure ~\ref{fig:Salinity_temp_data}).

We see in the bottom row of Figure ~\ref{fig:heatmap} where no noise is applied, that there is a sharp threshold where tipping starts to occur after the model has been forced strongly enough and for long enough ($u03 \ 100 \ yrs$). We conclude that this is bifurcation-induced tipping.

In the 
$u03 \ 100 \ yrs$ scenario, we unexpectedly see a lower probability of AMOC collapse at higher noise levels. We argue that this is because the noise forcing can push the model away from a tipping point, that is the noise provides a route for the AMOC to recover. We cannot have a tipping probability higher than unity, as in the no-noise scenario for this forcing, so the higher noise level results in a reduced probability of AMOC collapse.

In the $u01 \ 1000 \ yrs$ scenario, we see a significantly higher probability of collapse for high noise levels, than e.g. the $u01 \ 100 \ yrs$ scenario. 
In $u01 \ 1000 \ yrs$, the system is being forced closer to its tipping threshold (bifurcation point) for a long time, allowing more chance for the noise to induce a collapse.
We conclude that a noise-induced collapse over this time scale is most likely in scenarios with a low, long forcing and high noise levels, or a short, strong forcing with medium noise levels.
\begin{figure}
    \centering
    \begin{subfigure}{0.3\textwidth}
        \includegraphics[width=\textwidth]{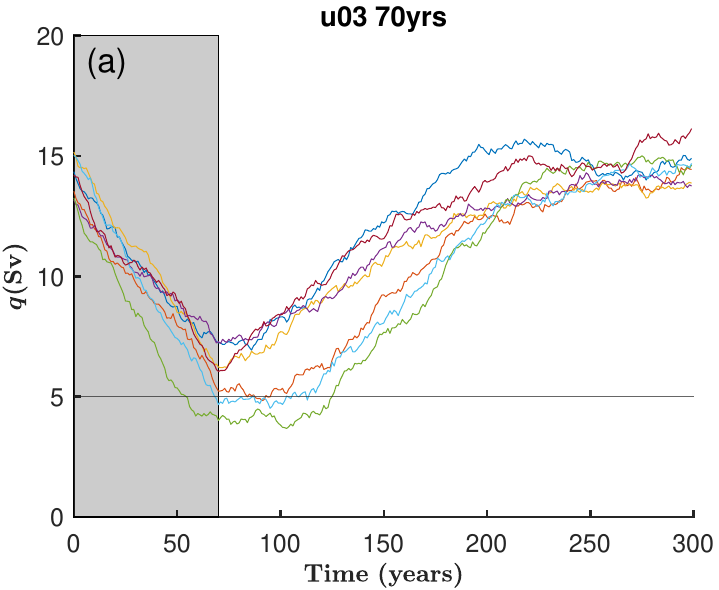}
    \end{subfigure}
    \hfill
    \begin{subfigure}{0.3\textwidth}
        \includegraphics[width=\textwidth]{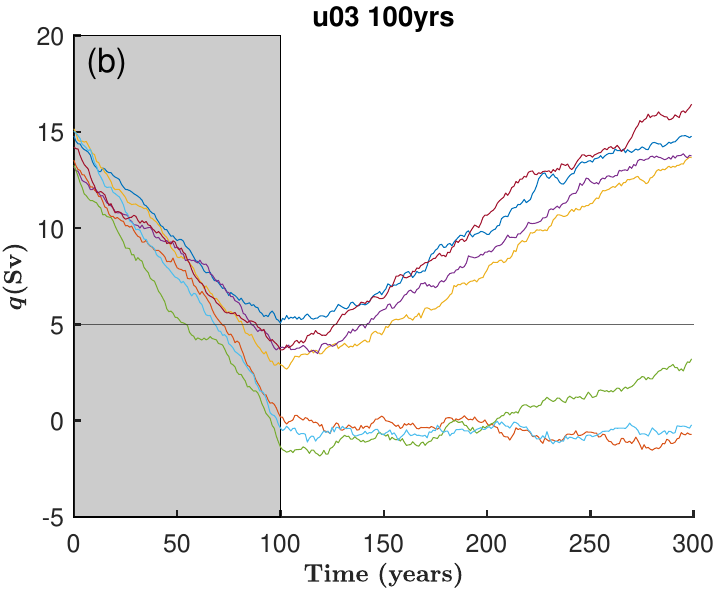}
    \end{subfigure}
    \hfill
    \begin{subfigure}{0.3\textwidth}
        \includegraphics[width=\textwidth]{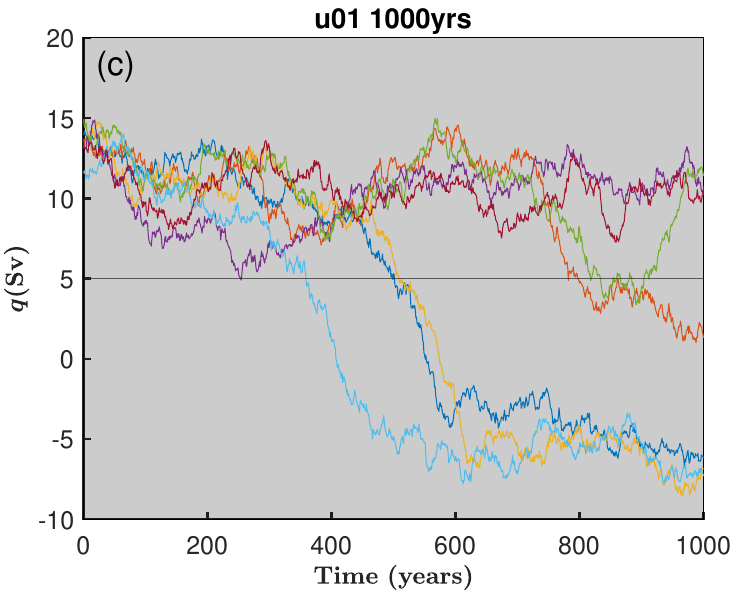}
    \end{subfigure}

    \caption{A selection of $q$ timeseries for runs of the non-linear three-box model with described noise and freshwater forcing. Each panel shows a handful of noise realisations from the 1000 ensemble Monte Carlo runs from Figure~\ref{fig:heatmap}. The $5 \ \rm{Sv}$ threshold of collapse is shown by the black line and forcing applied in the grey shaded region.
    (a) Trajectories forced with $u03\ 70 \ yrs$ and {\tt 5xHadGEM3MM} noise level.  (b) Equivalent trajectories for {\tt 5xHadGEM3MM} noise level and $u03 \ 100 \ yrs$ forcing. (c) Trajectories under $u01 \ 1000 \ yrs$ forcing, {\tt 10xHadGEM3MM} noise level. Note we show $1000$ years for panel (c), and $300$ years in panels (a) and (b).}
    \label{fig:tipping-trajectories}
\end{figure}

Figure~\ref{fig:tipping-trajectories} shows example trajectories of chosen noise and forcing scenarios from Figure~\ref{fig:heatmap}. In each panel of Figure~\ref{fig:tipping-trajectories}, one scenario of noise and forcing is plotted, with the grey region indicating when forcing is taking place. From the 1000 simulations of the box model used to obtain each panel of Figure~\ref{fig:heatmap}, $7$ typical trajectories are plotted to demonstrate different behaviors that the box model can exhibit. 

Figure~\ref{fig:tipping-trajectories}(a) shows a selection of trajectories 
for the $u03\ 70\ yrs$ scenario, with {\tt 5xHadGEM3MM} noise level. 
Only a small number ($\sim14\%$, Figure~\ref{fig:heatmap}) of trajectories drop below the $5 \ \rm{Sv}$ threshold for collapse.  For this scenario, the runs that drop below this value, appear to stay in a weakened state for longer than the runs that do not, which recover as soon as the forcing is switched off (e.g: the green line stays weak for several decades after hosing is removed). Since often GCMs are run for only around $150$ years \parencite{Jackson2023}, this result suggests that running GCMs for longer is necessary to assess the bi-stability and long-term fate of the AMOC.

Figure~\ref{fig:tipping-trajectories}(b) shows some trajectories for {\tt 5xHadGEM3MM} noise level and $u03\ 100\ yrs$ forcing. We see that many trajectories fall below the $5 \ \rm{Sv}$ threshold, but some continue to weaken below $0 \ \rm{Sv}$ to fully collapse, akin to a quasi-irreversible tipping. Unexpectedly, one realisation spontaneously recovers around 200 years, long after the forcing has been switched off. Similar behaviour has also been seen in GCMs \parencite{Stouffer2003}. 
Thus, we argue that a noise fluctuation can spontaneously recover the AMOC, even from a fully collapsed state.
 
Finally, for lower hosing levels, such as the $u01 \ 1000 \ yrs$ scenario in Figure~\ref{fig:tipping-trajectories}(c), we see that higher noise levels ({\tt 10xHadGEM3MM}) are able to collapse the AMOC. These fluctuations cause some trajectories to collapse to the $5 \ \rm{Sv}$ threshold, and many continue to weaken below $0 \ \rm{Sv}$ and fully collapse. In any scenario, a long enough run will lead to a noise-induced collapse, due to the nature of Gaussian white noise. In this case, considering $u01 \ 1000 \ yrs$, the model must be forced slightly ($0.1 \ \rm{Sv}$) for a noise-induced collapse to have a non-negligible probability within this time frame.

We note that the precise collapse times are unpredictable in this scenario: they only depend on the noise realisation and not on the forcing profile. This behaviour appears to be similar to that seen in a GCM (with CO${}_2$ rather than freshwater forcing) by \textcite{Romanou2023}.

\section{Conclusions}
\label{sec:conclusions}

In this study we find that the estimated noise for pre-industrial Control runs for several CMIP6 models is very small, and is unlikely to collapse the AMOC without significant freshwater forcing, beyond that expected with modern-day climate change.
Our findings progress beyond the CMIP5 study of \textcite{Castellana2019} to consider a selection of CMIP6 GCMs. We similarly find that the estimated noise from pre-industrial states is small and is not enough by itself to tip the AMOC on policy-relevant time scales, in a simple nonlinear box model. A freshwater forcing that shifts the AMOC closer to a bifurcation point 
is needed to produce noise-induced AMOC collapse.  
It is notable that the experiment with a sustained lower level of forcing ($u01\ 1000\ yrs$) is more likely to collapse than the high, short forcing scenario ($u03\ 50\ yrs$). We suggest this is because the system must be pushed near to its bifurcation point for a longer time for stochastic terms to induce a collapse. The box model has enough resilience to recover from the stronger forcing before the noise takes effect in the $u03\ 50\ yrs$ scenario. The sustained weakening and spontaneous recovery of the AMOC in the scenarios shown in Figures \ref{fig:tipping-trajectories}a and \ref{fig:tipping-trajectories}b would certainly have important climatic impacts if the AMOC were to weaken by this amount \parencite{Liu2020}.

The noise inferred from the CMIP6 models examined suggests only a minor role for internal variability in inducing an AMOC collapse. However, observations and historical simulations suggest that real-world variability may be substantially greater than that found in these models, and we note that recently \textcite{Romanou2023} have reported noise-induced AMOC collapse in a GCM. We therefore increase the variability beyond the GCM estimates, to understand the potential risks of a `real-world' noise on the AMOC, and find that this has a large impact on the probability of collapse. Our results highlight that better quantification of decadal variability of the AMOC is essential to understanding its potentially decisive role in AMOC tipping. The results for the $u03 \ 100 \ yrs $ scenario, where forcing was removed after $100$ years, are relevant for risks under climate overshoot scenarios, showing that in some cases of overshooting forcing, internal variability even may act to {\it reduce} the overall probability of collapse.

Future work could explore whether the processes driving decadal variability are comparable between CMIP6 models, and how the amplitude of this variability differs from observations.
Further hosing scenarios and scenarios driven by greenhouse gases/warming could also be considered, including overshoot scenarios relevant to safe climate mitigation pathways, where presence or absence of tipping may depend not only on the forcing but on the individual realization of the variability \cite{lohmann2024predictability}. 
It would also be interesting to consider state-dependant noise (e.g. noise that depends on the strength of the AMOC).

Our results confirm that the inclusion of variability can increase (or decrease) the risk of an AMOC collapse and will certainly contribute to the uncertainty of AMOC projections. Reducing the forcing decreases the risk of AMOC collapse, but internal decadal variability will be present independent of anthropogenic effects. This suggests that it remains important to include realistic models of decadal variability in evaluating mitigation pathways. 

\section*{Acknowledgements}

For the purpose of open access, the authors have applied a Creative Commons Attribution (CC BY) licence to any Author Accepted Manuscript version arising from this submission. RC thanks EPSRC for support via an EPSRC CASE PhD studentship with the Met Office and the University of Exeter grant agreement number EP/T518049/1. PA thanks EPSRC for support via EP/T018178/1. RW and JB were supported by the Met Office Hadley Centre Climate Programme funded by BEIS, and by the European Union’s Horizon 2020 research and innovation programme under grant agreement no. 820970 (TiPES project). This paper is TiPES contribution number 275.
We gratefully acknowledge the following for insightful discussions: Laura Jackson, Henk Dijkstra, Paul Ritchie, Lee de Mora, Swinda Falkena and Johannes Lohmann.
We acknowledge the World Climate Research Programme for its coordination of CMIP6, the climate modelling groups for producing and making available their model output, the Earth System Grid Federation (ESGF) for archiving the data and providing access, and the multiple funding agencies who support CMIP6 and ESGF.

\section*{Data Availability Statement}

Datasets for this research are available in \textcite{Dataset}.

\newpage
\printbibliography

\appendix

\newpage

\section*{Supplementary Information}

\renewcommand{\thetable}{S\arabic{table}}  
\renewcommand{\thefigure}{S\arabic{figure}}

\section*{\bf{S1} Details of CMIP6 runs used}
\label{app:model details}

Table \ref{tab:model_details} lists the details of the CMIP6 runs considered. We choose these models as a subset of the models used in \textcite{Jackson2023} to be able to compare the results of the box models to the GCM runs under the same freshwater forcing scenario. There was some difficulty accessing the required data in JASMIN (Centre for Environmental Data Analysis archive) , as consequently these models were used since these are the once for which the data averaging and extraction was successful. We begin with two resolutions of {\tt HadGEM3} and add two additional models with different resolutions and ocean components for comparison.

All data (Figure 1 and 3, main article) was taken from CMIP6, with some of the following details;
MIP: Omon, Experiment: piControl, Ensemble: r1i1p1f1, Grid: gn.

The {\tt CanESM5} and {\tt MPI} data was averaged from JASMIN using {\tt esmvaltool} \parencite{esmvaltool}. The Northern box was extracted using a shapefile, created in ArcGISpro. This allows us to define an accurate shape with multiple vertices. For the Trop. box, we defined maximum and minimum values for the latitude, longitude and depth of the region. The {\tt HadGEM3} data was extracted from the model at the Met Office using a land-sea mask file.

To verify these extraction methods, the extracted time series for {\tt HadGEM3} (-LL and -MM) models from JASMIN using shapefiles were compared with the land-sea mask method implemented at the Met Office and found to give good agreement. All time series do not include the Mediterranean sea.

\begin{table}    
    \centering
    \begin{tabular}{p{2cm} p{2.5cm} p{2.5cm} p{2.5cm} p{2.5cm}}
    \hline
    Model  & {\tt HadGEM3LL} & {\tt HadGEM3MM} & {\tt CanESM5} & {\tt MPI-ESM1-2-LR}  \\
\hline
Start Year & 1850 & 1850 & 5400 & 1950 \\
End Year & 2350 & 2350 & 5699 & 2249 \\
Ocean Component & NEMO & NEMO & NEMO & MPIOM\\
Vertical Resolution & 75 levels & 75 levels & 45 levels & 40 levels\\
Horizontal resolution & ORCA1 ($1 ^{\circ}$) & ORCA025 ($0.25^{\circ}$) & ORCA1 ($1^{\circ})$ & $1.5^{\circ}$\\
Reference & \cite{Kuhlbrodt2018}, \cite{Williams2018}, \cite{storkey2018} & (same as low resolution model) & \cite{Swart2019} & \cite{muller2018}, \cite{jungclaus2013}\\
\hline
\end{tabular}
\caption{Details of CMIP6 pre-industrial (pi) control experimental runs analysed in this paper.}
    \label{tab:model_details}
\end{table}

\section*{\bf{S2} Summary Statistics of CMIP6 runs}
\label{app:CMIP6_stats}

In Table \ref{tab:variability} we present summary statistics of the time series of salinity, temperature and AMOC strength for the CMIP6 models.
$\overline{S_{\rm{Nor.}}}$, $\overline{S_{\rm{Trop.}}}$ are the time-mean values for each run, while the (co)variances are $\Sigma_{\rm{XY}}= \overline{(S_{\rm{X}}-\overline{S_{\rm{Y}}})(S_{\rm{X}}-\overline{S_{\rm{Y}}})}
$ where $X,Y\in\{\rm{Nor.,Trop.}\}$. Equivalent results for temperature and AMOC strength are denoted by $T$ and $q$.

\begin{table}
\centering
\begin{tabular}{l c c c c}
\hline
Model  & {\tt HadGEM3LL} & {\tt HadGEM3MM} & {\tt CanESM5} & {\tt MPI-ESM1-2-LR}  \\
\hline
$\overline{S_{\rm{Nor.}}}$ (psu) & $35.0795$ &$35.0806$&$34.8225$&$35.0606$\\ 
$\overline{S_{\rm{Trop.}}}$ (psu) &$35.7997$&$35.6018$&$35.5637$&$35.3735$\\
\hline
$\Sigma_{S_{\rm{Nor.Nor.}}} (\mbox{[psu]}^2 \times 10^{-4})$&$0.3250$&$0.5412$ &$0.8252$&$1.249$\\ 
$\Sigma_{S_{\rm{Trop.Trop.}}} (\mbox{[psu]}^2\times 10^{-4})$ &$0.8074$&$0.7912$&$0.2547$&$0.235$\\ 
$\Sigma_{S_{\rm{Nor.Trop.}}} (\mbox{[psu]}^2\times 10^{-4})$ & $-0.2031$&$-0.1779$&$-0.2329$&$0.188$\\
\hline
$\overline{T_{\rm{Nor.}}} ~(^{\circ}{\rm{C}})$  & $3.2995$ & $3.2995$ & $4.4009$ & $6.1206$ \\
$\overline{T_{\rm{Trop.}}} ~(^{\circ}{\rm{C}})$ & $14.7207$ & $13.7534$ & $12.6184$ & $12.3430$ \\
\hline
$\Sigma_{T_{\rm{Nor.Nor.}}} ~([^{\circ}{\rm{C}}]^{2})$ & $0.0010$ & $0.0010$ & $0.0017$ & $0.0035$ \\
$\Sigma_{T_{\rm{Trop.Trop.}}} ~([^{\circ}{\rm{C}}]^{2})$ & $0.0023$ & $0.0037$ & $0.0008$ & $0.0003$ \\
$\Sigma_{T_{\rm{Nor.Trop.}}} ~([^{\circ}{\rm{C}}]^{2})$ & $0.0006$ & $0.0010$ & $-0.0007$ & $-0.0003$ \\
\hline
$\bar{q} ~\rm{(Sv)}$ &$13.3581$&$15.3115$&$17.6939$&$18.9147$\\ 
$\Sigma_{q} ~\rm{([Sv]^2)}$ &$0.1644$&$0.1497$&$1.4941$&$0.2674$\\
\hline
\end{tabular}
\caption{Means (indicated by bar notation) and covariance ($\Sigma$) of the regional volume averaged decadal mean salinity, temperature and AMOC timeseries presented in Figure 1 and 3, for the $300$ year CMIP6 piControl runs considered, details in Table \ref{tab:model_details}.}
\label{tab:variability}
\end{table}

\section*{\bf{S3} Equations and Parameters for the three-box model}
\label{app:3box_parameters}

We note that \textcite{Alkhayuon19} simplify the 5-box model of \textcite{wood19} to consider a closed system of equations for the Northern and Tropical Atlantic boxes, depending on the AMOC flow $q$ given by Eq.(1). The boxes considered correspond to the following; `S' is the Southern Ocean body of water near the surface; `Trop.' is the Tropical Atlantic from the surface to the thermocline; `Nor.' is the North Atlantic, where colder water sinks downwards forming the North Atlantic Deep Water (NADW); `B'  represents the Bottom waters and `IP' is the Indo-Pacific box from the surface to the thermocline. We make a reduction to three boxes by considering the `S' and `B' boxes to be constant and solving for the indo-pacific box.
For $q\ge0$ this is
\begin{subequations}
\begin{flalign}
    V_{\rm{Nor.}} \frac{dS_{\rm{Nor.}}}{dt} =&q (S_{\rm{Trop.}} - S_{\rm{Nor.}}) + K_{\rm{Nor.}}(S_{\rm{Trop.}} - S_{\rm{Nor.}}) - F_{\rm{Nor.}} S_{0},\\
    V_{\rm{Trop.}} \frac{dS_{\rm{Trop.}}}{dt} = & q [\gamma S_{\rm{S}} + (1-\gamma)S_{\rm{IP}} - S_{\rm{Trop.}}] + K_{\rm{S}}(S_{\rm{S}} - S_{\rm{Trop.}}) \\
    &+ K_{\rm{Nor.}}(S_{\rm{Nor.}} - S_{\rm{Trop.}}) - F_{\rm{Trop.}} S_{0},\nonumber
\end{flalign}
\label{eq:3box_on}
\end{subequations}
while for $q<0$ this is
\begin{subequations}
\begin{flalign}
   V_{\rm{Nor.}} \frac{dS_{\rm{Nor.}}}{dt} = &|q| (S_{\rm{B}} - S_{\rm{Nor.}}) + K_{\rm{Nor.}}(S_{\rm{Trop.}} - S_{\rm{Nor.}}) - F_{\rm{Nor.}} S_{0},\\
    V_{\rm{Trop.}} \frac{dS_{\rm{Trop.}}}{dt} = &|q|(S_{\rm{Nor.}} - S_{\rm{Trop.}}) + K_{\rm{S}}(S_{\rm{S}} - S_{\rm{Trop.}}) \\
   & + K_{\rm{Nor.}}(S_{\rm{Nor.}} - S_{\rm{Trop.}}) - F_{\rm{Trop.}} S_{0}.\nonumber
\end{flalign}
\label{eq:3box_off}
\end{subequations}
Global conservation of freshwater flux implies that the total salt content $C$ is constant:
\begin{equation}
    C = V_{\rm{Nor.}}S_{\rm{Nor.}}+V_{\rm{Trop.}}S_{\rm{Trop.}}+V_{\rm{S}}S_{\rm{S}}+V_{\rm{IP}}S_{\rm{IP}}+V_{\rm{B}}S_{\rm{B}}.
    \label{eq:salt}
\end{equation}
These constraints allow us to eliminate one of the differential equation when looking at solutions of the system. When we choose the $S$ and $B$ boxes to be constant, the conservation of salt given in Equation \ref{eq:salt} allows us to solve for $S_{\rm{IP}}$;
\begin{equation}
    S_{\rm{IP}} = \frac{(C - V_{\rm{Nor.}}S_{\rm{Nor.}} - V_{\rm{Trop.}}S_{\rm{Trop.}} - V_{\rm{S}}S_{\rm{S}} - V_{\rm{B}}S_{\rm{B}})}{V_{\rm{IP}}}.
    \label{eq:SIP}
\end{equation}
The parameters used are given in Table~\ref{tab:params}, which are based on the default parameters in \cite{Alkhayuon19}, updated for {\tt HadGEM3}.

For the stochastic simulation note that $\bm{Q} = \bm{B} \bm{B}^\top$ so that  $\bm{B}$ is a `pseudo square root' of the covariance matrix \parencite[Proposition 11.31]{Breiman92}. Note that pseudo square roots are not unique. However there exists a unique lower triangular matrix $\bm{B}$ with positive diagonal such that $\bm{B}$ is a pseudo square root of $\bm{Q}$. This matrix is a Cholesky decomposition of $\bm{Q}$. We assume that the noise covariance must be positive definite so that a unique pseudo square root can be found. 
We parametrise $\bm{Q}$ by
 \begin{equation}
 \bm{Q}_{11}=\exp \sigma_1,~~\bm{Q}_{22}=\exp \sigma_2,~~\bm{Q}_{12}=\bm{Q}_{21}=(\exp \sigma_1+\exp \sigma_2)\tanh \sigma_2,
 \end{equation}
which gives a parametrisation of the the space of positive definite $\bm{Q}$ in terms of the real parameters $\sigma_{1,2,3}$.

\begin{figure}
    \centering
    \includegraphics[width = 0.8\linewidth]{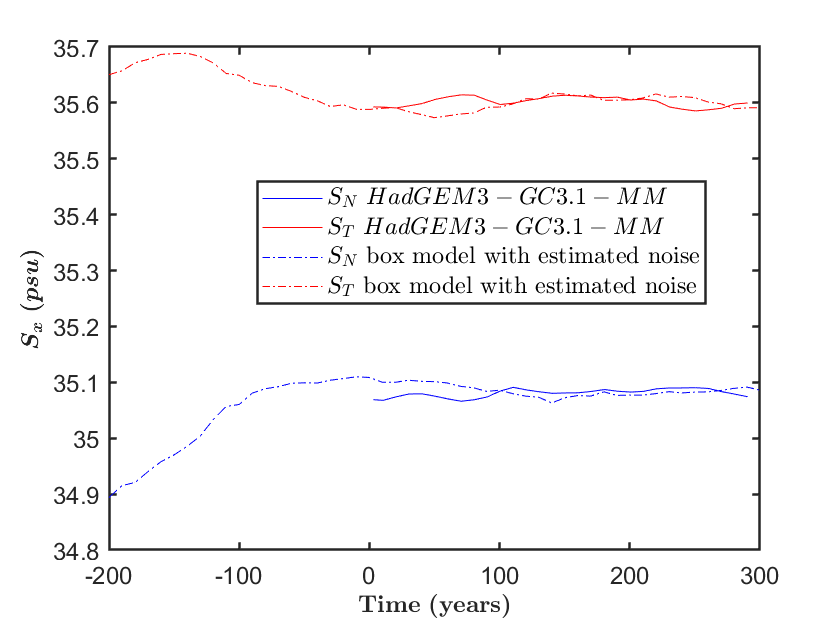}
    \caption{A reconstruction of the three box model salinities with noise estimates is shown in dashed lines, using {\tt HadGEM3MM} model parameters. The model is spun up for 200 years to allow for differences in the initial values of the salinity. Also plotted is the {\tt HadGEM3MM} `data' extracted directly from the model for the ocean regions. The Northern box is shown in blue and the Tropical box shown in red.}
    \label{fig:HadGEM3MM_reconstruction}
\end{figure}

We include in Figure~\ref{fig:HadGEM3MM_reconstruction} an example simulation of the three box model with {\tt HadGEM3MM} box model calibration (Table \ref{tab:params}) and noise estimate, as decomposed from the covariance $\bm{Q}$ and presented in Table 1. The three-box model simulation is given a short spin up. Also plotted is the volume averaged salinity time series for {\tt HadGEM3MM} for both the $Nor.$ and $Trop.$ boxes with no forcing. This Figure shows very good agreement between the re-calibrated box model with estimated noise and optimisation, and the GCM results.

\begin{table}
\centering
{\small
\begin{tabular}{l c c}
\hline
Parameter  & HadGEM3LL & HadGEM3MM \\
\hline
$V_{\rm{Nor.}} (m^3 \times 10^{16})$ &  4.633 & 4.192\\
$V_{\rm{Trop.}} (m^3 \times 10^{16})$ & 4.358 &  4.191\\ 
$V_{\rm{S}} (m^3 \times 10^{16})$ & 11.13 & 13.26\\ 
$V_{\rm{IP}} (m^3 \times 10^{16})$ & 18.46 & 16.95\\ 
$V_{\rm{B}} (m^3 \times 10^{16})$ & 94.98 & 96.76\\  
\hline
$F_{\rm{Nor.}}^0 (\rm{Sv})$ & 0.5296 & 0.5996\\ 
$F_{\rm{Trop.}}^0 (\rm{Sv})$ &  -0.8022 & -0.7527\\ 
$F_{\rm{S}}^0 (\rm{Sv})$ & 0.9088 & 1.126\\
$F_{\rm{B}} (\rm{Sv})$  & 0 & 0\\ 
\hline
$T_{\rm{S}} (^o C)$ & 5.831 & 5.349\\
$T_{\rm{0}} (^o C)$ & 5.33 & 4.514\\ 
\hline
$K_{\rm{Nor.}} (\rm{Sv})$ & 6.58 & 4.73\\
$K_{\rm{S}} (\rm{Sv})$  & 2.68 &  7.68\\
$K_{\rm{IP}} (\rm{Sv})$ & 106.09 & 123.49\\ 
\hline
$\alpha$ & 0.12 & 0.12\\
$\beta$ & 790 & 790 \\
$\eta (\rm{Sv})$ & 4.821 & 6.211\\
$\gamma$ & 0.58 & 0.58\\
$\lambda (m^6 kg^{-1} s^{-1} \times 10^{7})$ & 1.69 & 2.328\\
$\mu (^o C m^{-3}s \times 10^{-8})$ & 0 & 0\\ 
\hline
$A_{\rm{Nor.}}$ & 0.9802 & 0.9841\\
$A_{\rm{Trop.}}$ & -0.1752 & -0.1853\\
$A_{\rm{S}}$ & -0.2045 & -0.1742\\
$A_{\rm{IP}}$ & -0.6005 & -0.6245\\
\hline
\end{tabular}
}
  \caption{Parameters for three-box model as extracted directly from {\tt HadGEM3} to calibrate the three-box model to the piControl runs, based on the years 2000-2050. Note that the values of $\alpha$ and $\beta$ are taken from previous estimates and have not been updated. $F_{\rm{IP}}$ provides a balance for the freshwater fluxes to sum to zero. 
  }
\label{tab:params}
\end{table}

\section*{\bf{S4} Maximum likelihood estimation of noise covariance and parameters for the three box model}
\label{app:mle}

We minimise the following function to estimate parameters for the non-linear model (Equation 2);
\begin{equation}
\begin{split}
    l (\bm{\sigma}) = 
    \sum_{k=0}^{T-1}  \frac{1}{2} \Bigg[ {\rm{log}} | 2\pi  \bm{Q(\sigma) } \Delta t |  + (\bm{x}_{\rm{k+1}} - \bm{x}_{\rm{k}} - \bm{f}_{\rm k} \Delta t) ^\top  ( \bm{Q(\sigma)} \Delta t )^{-1} (\bm{x}_{\rm{k+1}} - \bm{x}_{\rm{k}} - \bm{f}_{\rm k} \Delta t) \Bigg],
    \label{eq:negloglikelihood}
\end{split}
\end{equation}
where $\bm{f}_{\rm k}=\bm{f}(\bm{x}_{\rm{k}},t_{\rm{k}};\bm{\sigma})$ and $\bm{x}_{\rm k}=\bm{x}(t_{\rm{k}})$.
Equation~\ref{eq:negloglikelihood} is a log-likelihood approximation \parencite{sarkka19} for the nonlinear stochastic model (Equation (2)) where $\bm{x}_{\rm{k}}$ corresponds to the `data' point at each time step, $\bm{f}_{\rm{k}}$ is the model evaluated at the equivalent time step, $\bm{\sigma}$ are parameters to be estimated, $\bm{Q}(\bm{\sigma})$ is the noise covariance (also parametrised by $\bm{\sigma}$). 
Equation \ref{eq:negloglikelihood} finds the bias between the `data' and the model simulation, and works to minimise this bias by optimising the estimated parameters in the vector $\bm{\sigma}$. The driving noise covariance $\bm{Q}$ can be estimated more directly than $\bm{B}$ in Equation (2).
We parameterise $\bm{Q}$ to be positive definite in terms of real parameters $\sigma_{1,2,3}$ so that a pseudo-square root $\bm{B}$ is then found via Cholesky decomposition since $\bm{Q} = \bm{B} \bm{B}^\top$. This will typically have multiple solutions and so we can assume that $\bm{B}$ is lower diagonal with positive entries to find a unique solution. We minimize $l(\bm{\sigma})$ using the Matlab function {\tt fminsearch}. The optimisation parameters of this function were set to achieve a given tolerance, that is when the likelihood function changes by less than $10^{-6}$.

The log likelihood function (Equation~\ref{eq:negloglikelihood}) can also be used to optimize the freshwater fluxes; this allows us to fit the mean salinities from the data to the model at the same time as determining the noise. To this end, we introduce parameters $\sigma_{4,5}$ and set the freshwater fluxes:
\begin{equation}
 F_{\rm{Nor.}} = F_{\rm{Nor.}}^0 + 10^6\delta_{F_{SNor.}},~~
 F_{\rm{Trop.}} = F_{\rm{Trop.}}^0 + 10^6\delta_{F_{Srop.}},~~
 F_S = F_S^0 - 10^6(\delta_{F_{SNor.}} + \delta_{F_{STrop.}}),
\end{equation}
where $F_{\rm{Nor.}}^0$, $F_{\rm{Trop.}}^0$ and $F_S^0$ are the original values for the parameters; note the parameters are chosen to ensure that the fluxes are still balanced. $\delta_{F_{SNor.}}$ and $\delta_{F_{STrop.}}$ are the last two entries in the vector ($\bm{\sigma}$), $\sigma_4$ and $\sigma_5$ respectively. They represent the amount that the parameters $F_{\rm{Nor.}}$ and $F_{\rm{Trop.}}$ are changed by when we optimise them. The subscript e.g: $SNor.$ indicates water moving from the $S$ box to the Nor. box.
Note there is a consistent bias $\delta_{F_{SNor}}<0$ corresponding to $S_{\rm{Nor.}}$ being consistently less salty for the standard parameters in the 3-box model than in the CMIP6 models.  Interestingly, $\bm{Q}_{21}$ (and correspondingly $\bm{B}_{21}$) are indicating a weakly negative covariance between regions.

The results of the noise estimation are shown in Table 1 (noise amplitudes) and \ref{tab:Qestimated} (noise covariance and freshwater flux optimisation). Here we display the estimates of the noise covariance, $\bm{Q}$, which is the direct output of the log-likelihood. 

\begin{table}
\centering
\begin{tabular}{l c c c c}
\hline
Model  & {\tt HadGEM3LL} & {\tt HadGEM3MM} & {\tt CanESM5} & {\tt MPI}  \\
\hline
$\bm{Q}_{11} (\times 10^{-11})$ & 0.1135 & 0.1595 & 1.4900 & 7.545\\ 
$\bm{Q}_{21}=\bm{Q}_{12} (\times 10^{-11})$ & -0.0186 & -0.1098 & -0.6100 & -0.3850\\ 
$\bm{Q}_{22} (\times 10^{-11})$ & 0.1635 & 0.1939 & 1.0290 & 1.2750\\ \hline
$\delta_{F_{SNor.}}$ (Sv) & -0.1538 & -0.3197 & -0.1984 & -0.3276\\ 
$\delta_{F_{STrop.}}$ (Sv) & 0.0686 & -0.0393 & 0.3122 & 0.3464\\
\hline
\end{tabular}
\caption{Estimated noise covariance and optimised freshwater flux values of the nonlinear 3-box model (Equation (2)) estimated using the approximated  log likelihood function from \ref{app:mle} for the decadal CMIP6 data shown in Figure 3. $\bm{Q}$ indicates the estimated noise covariance where the subscripts indicate matrix component (1 = N, 2 = T), in SI units. $\delta_{F}$ are the optimisation values for the freshwater flux, the initial values of $F$ are listed in \ref{app:3box_parameters}}
\label{tab:Qestimated}
\end{table}

\end{document}